\documentstyle[multicol,aps,prl]{revtex} 

\renewcommand{\narrowtext}{\begin{multicols}{2} \global\columnwidth20.5pc}
\def\al{\alpha}
\def\be{\beta}
\def\ga{\gamma}
\def\de{\delta}

\def\ve{\varepsilon}

\def\th{\theta}

\def\ka{\kappa}

\def\si{\sigma}

\def\ta{\tau}

\def\ph{\phi}

\def\ch{\chi}
\def\ps{\psi}
\def\om{\omega}

\def\Om{\Omega}

\def\cl{{\cal L}}

\def\fr#1#2{{{#1} \over {#2}}}

\def\half{{\textstyle{1\over 2}}}
\def\frac#1#2{{\textstyle{{#1}\over {#2}}}}

\def\lsim{\mathrel{\rlap{\lower4pt\hbox{\hskip1pt$\sim$}}
    \raise1pt\hbox{$<$}}}
\def\gsim{\mathrel{\rlap{\lower4pt\hbox{\hskip1pt$\sim$}}
    \raise1pt\hbox{$>$}}}
\def\sqr#1#2{{\vcenter{\vbox{\hrule height.#2pt
         \hbox{\vrule width.#2pt height#1pt \kern#1pt
         \vrule width.#2pt}
         \hrule height.#2pt}}}}

\def\lrpartial{\raise 1pt\hbox{$\stackrel\leftrightarrow\partial$}}

\def\lrdnu{\stackrel{\leftrightarrow}{D^\nu}}

\newcommand{\beq}{\begin{equation}}
\newcommand{\eeq}{\end{equation}}
\newcommand{\bea}{\begin{eqnarray}}
\newcommand{\eea}{\end{eqnarray}}
\newcommand{\rf}[1]{(\ref{#1})}
 
\begin{document}

\title{Lorentz and CPT tests with spin-polarized solids}

\author{Robert Bluhm$^a$ and V.\ Alan Kosteleck\'y$^b$} 

\address{$^a$Physics Department, Colby College, Waterville, ME 04901} 
\address{$^b$Physics Department, Indiana University, Bloomington, IN 47405} 

\date{IUHET 414, COLBY-99-05;
accepted for publication in Phys.\ Rev.\ Lett.}

\maketitle

\begin{abstract}
Experiments using macroscopic samples of spin-polarized matter
offer exceptional sensitivity to Lorentz and CPT violation
in the electron sector.
Data from existing experiments with a spin-polarized torsion pendulum 
provide sensitivity in this sector 
rivaling that of all other existing experiments
and could reveal spontaneous violation of Lorentz symmetry 
at the Planck scale.
\end{abstract}

\narrowtext

The standard model of particle physics is 
invariant under Lorentz and CPT transformations
\cite{sachs,cpt98}.
However,
in an underlying theory combining the standard model with gravity,
Lorentz and CPT symmetry might be spontaneously broken
\cite{kskp}.
Small low-energy signals of Lorentz and CPT breaking 
might then be detectable in high-precision tests.
The dimensionless suppression factor for such effects
would be the ratio of a low-energy scale to the Planck scale,
perhaps combined with dimensionless coupling constants.

Many high-precision tests of Lorentz and CPT symmetry 
in matter are spectroscopic in the sense that they 
involve measuring or monitoring frequencies
associated with particles or atoms.
Examples include 
comparative studies of anomaly and cyclotron 
frequencies of trapped particles and antiparticles
\cite{penningtests,bkr9798}
and clock-comparison experiments 
\cite{cctests,kl99a}.
These are often regarded as the sharpest tests of Lorentz 
and CPT symmetry in matter.
In the electron sector,
for instance,
it is possible with some theoretical assumptions
to bound frequency differences due to Lorentz and CPT violation
at the level of about $10^{-27}$ GeV.

In this work,
we examine an alternative class of experiments 
involving studies of the behavior of macroscopic solid matter.
The idea is to search for Lorentz- and CPT-violating spin couplings
using materials with a net spin polarization,
produced by the combined effects of many electrons.
We show that a particular type of experiment is presently capable 
of testing Lorentz and CPT symmetry 
in the electron sector with a precision rivaling 
that of spectroscopic experiments.

A variety of experiments using spin-polarized matter exist. 
They include,
for example,
studies of torques on a spin-polarized torsion pendulum
\cite{phillips87,ritter93,ega98,harris98}
and measurements of the induced magnetization 
in a paramagnetic salt using a dc SQUID 
\cite{ni99}.
Except for the experiment in Ref.\ \cite{phillips87},
which was designed to test spatial isotropy, 
the primary motivation of these experiments has been to search
for anomalous spin couplings associated with spin-gravitational 
effects and axion couplings
\cite{torpend},
for which recently attained sensitivities exceed 
those of spectroscopic searches 
\cite{spectests}.

To investigate the sensitivity 
to Lorentz and CPT violation
of experiments with spin-polarized matter,
we use a general standard-model extension
\cite{ck}
describing effects arising in any fundamental theory 
in which spontaneous Lorentz and CPT breaking occurs.
The theory provides a consistent microscopic description  
of these effects in the context of an otherwise conventional 
renormalizable quantum gauge field theory.
In addition to the trapped-particle and clock-comparison tests
mentioned above
\cite{penningtests,bkr9798,cctests,kl99a},
the theory has been applied to
spectroscopic comparisons of hydrogen and antihydrogen
\cite{bkr99},
experiments with muons
\cite{bkl99},
tests with neutral-meson oscillations
\cite{mesontests,ckpv},
searches for cosmic birefringence
\cite{cfj,ck,jk99},
measurements of the baryon asymmetry 
\cite{bckp},
and observations of cosmic rays
\cite{cg}.

The macroscopic samples of spin-polarized matter
used in the experiments of interest here,
such as a spin-polarized torsion pendulum or a paramagnetic salt crystal,
have a large net electron spin and negligible net nuclear spin.
According to the above general theoretical framework,
a sample of this type experiences an effective potential
arising from the coupling of the electron angular momenta
to spacetime-independent background tensors
generating the Lorentz and CPT violation.
The first step in determining this potential
is to extract an appropriate quantum-electrodynamics limit 
of the standard-model extension describing the
Lorentz- and CPT-violating effects on electrons.
In units with $\hbar = c = 1$,
the relevant perturbative Lorentz-violating lagrangian terms are 
\bea
\cl &=&
- a_{\mu}^e \, \bar \ps \ga^\mu \ps
- b_{\mu}^e \, \bar \ps \ga_5 \ga^\mu \ps
- \half H_{\mu\nu}^e \,
  \bar \ps \si^{\mu\nu} \ps
  \nonumber \\
&&
+ \half i c_{\mu\nu}^e \,
  \bar \ps \ga^\mu \lrdnu \ps
+ \half i d_{\mu\nu}^e \,
  \bar \ps \ga_5 \ga^\mu \lrdnu \ps
\quad ,
\label{lqed}
\eea
where $\ps$ denotes the electron field and
$iD_\mu \equiv i \partial_\mu - q A_\mu$ with
charge $q=-|e|$.
The five parameters 
$a_{\mu}^e$, $b_{\mu}^e$, $H_{\mu\nu}^e$,
$c_{\mu\nu}^e$, $d_{\mu\nu}^e$ 
govern the (small) magnitudes of the Lorentz violation,
with the CPT-odd terms being
associated with the first two.

The electrons in the spin-polarized materials are nonrelativistic.
The appropriate perturbative hamiltonian $\de h_n$ for the $n$th electron 
can be derived from the lagrangian \rf{lqed}
using established procedures involving field redefinitions
and a Foldy-Wouthuysen transformation
\cite{kl99b}.
The multiparticle perturbative hamiltonian $\de h$ describing 
leading-order Lorentz- and CPT-violating effects 
in the macroscopic spin-polarized material
can be obtained by summing over $n$.
Various physical properties can then be deduced from $\de h$.
For example,
energy-level shifts induced by Lorentz and CPT violation
can be found by taking expectation values
in an appropriate multiparticle quantum state.

Although the form of $\de h$ is lengthy in detail,
the dominant components relevant here 
can be shown to have the form
\beq
\de h\supset -\tilde b_j^e \sum_n \si_n^j
\quad . 
\label{h}
\eeq
This equation describes the coupling 
of the electron spins $\si_n^j$ to a combination 
$\tilde b_j^e $
of CPT-even and CPT-odd parameters for Lorentz violation
given by
\beq
\tilde b_j^e 
\equiv b_j^e - m d_{j0}^e - \half \ve_{jkl} H_{kl}^e
\quad .
\label{btilde}
\eeq
In these expressions,
Lorentz indices are separated into timelike and spacelike
cartesian components ($\mu = 0$ and $j = 1,2,3$),
and repeated indices are understood to be summed.
Other pieces of $\de h$ generate at most suppressed contributions
to the effective potential for spin-polarized matter.
For example,
although components involving the orbital angular momenta may appear, 
their expectation values and hence their contributions 
are suppressed by factors of order $\al^2 \simeq 5 \times 10^{-5}$ 
relative to those in Eq.\ \rf{h} and so can be disregarded.

The form of $\de h$ has some immediate implications for experiments
with macrosopic spin-polarized materials.
For example,
one type of experiment searches for anomalous spin-spin couplings 
by seeking effects when the relative orientation 
of two nearby spin-polarized masses is changed.
However,
$\de h$ contains no terms coupling electron spin to an external spin,
and so no signal for Lorentz or CPT violation
can be expected in experiments of this type.
Other types of experiment search for spin-monopole couplings,
with which Eq.\ \rf{h} is certainly compatible.
Nonetheless,
even these experiments 
are insensitive to the effects in Eq.\ \rf{h} 
unless the spin-polarized material is directly monitored.
For example,
the experiment of Ref.\ \cite{ritter93}
studies the behavior of an unpolarized torsion pendulum 
in the presence of an external spin-polarized mass
and therefore cannot detect couplings of the form \rf{h}. 
In contrast,
experiments studying the behavior 
of a spin-polarized torsion pendulum
\cite{phillips87,ega98,harris98}
or measuring changes in magnetization 
in a paramagnetic salt
\cite{ni99}
can be exquisitely sensitive to the couplings \rf{h}.

Consider first experiments with a spin-polarized torsion pendulum.
Choosing the direction $\hat z$ in the laboratory frame
as vertically upwards along the pendulum rotation axis,
the explicit expression for the perturbative contribution
to the potential energy of the pendulum is 
\beq
U(\ph) = 2 S \, 
\sqrt{(\tilde b_1^e)^2 + (\tilde b_2^e)^2} \cos\ph
\quad .
\label{U}
\eeq
Here,
$S$ is the net electron spin of the polarized pendulum,
and $\ph$ is the angle between the spin vector $S^j$
and the projection of the vector $\tilde b^{ej}$ 
on the $x$-$y$ plane.
The factor 
$\sqrt{(\tilde b_1^e)^2 + (\tilde b_2^e)^2}$
is the magnitude of this projection.

Experimental determination of the behavior of a spin-polarized pendulum
typically requires data collection over many hours.
During this time,
the sidereal rotation of the Earth 
changes the orientation of the laboratory-frame coordinates
relative to the background tensors
$a_{\mu}^e$, $b_{\mu}^e$, $H_{\mu\nu}^e$,
$c_{\mu\nu}^e$, $d_{\mu\nu}^e$.
In the laboratory frame,
the parameters $\tilde b^e_j$ in Eq.\ \rf{btilde}
therefore appear to be time dependent. 
To determine the corresponding time dependence of the potential $U$,
it is useful to work with quantities defined 
with respect to a nonrotating frame.
A suitable choice of basis $\{ \hat X, \hat Y, \hat Z \}$ for a
nonrotating frame can be introduced in terms of
celestial equatorial coordinates 
\cite{kl99a}.
With this choice,
the $\hat Z$ direction lies along the Earth's rotational north pole,
subtending an angle $\ch$ with the pendulum rotational axis $\hat z$.
The time dependence of the laboratory-frame components $\tilde b_j^e$
can then be displayed explicitly 
in terms of nonrotating-frame components as
\bea
\tilde b_1^e 
&=& 
\tilde b^e_X \cos \ch \cos \Om t
+ \tilde b^e_Y \cos \ch \sin \Om t - \tilde b^e_Z \sin \ch
\quad ,
\nonumber\\
\tilde b_2^e 
&=& 
- \tilde b^e_X \sin \Om t + \tilde b^e_Y \cos \Om t 
\quad ,
\nonumber\\
\tilde b_3^e 
&=& 
\tilde b^e_X \sin \ch \cos \Om t
+ \tilde b^e_Y \sin \ch \sin \Om t + \tilde b^e_Z \cos \ch
\quad .
\label{b2}
\eea
Here, 
the angular frequency $\Om$ 
is the Earth's sidereal (not solar) rotational frequency,
$\Om \simeq {2 \pi}/{({\rm 23 h \, 56 m})}$.

At present,
the spin-polarized torsion pendulum most sensitive
to Lorentz- and CPT-violating effects is the one used with 
the E\"ot-Wash II instrument at the University of Washington 
\cite{ega98,harris98}.
It has four stacked layers of toroidal magnets with 
alternating sections made of Alnico and SmCo,
producing a large net electron spin
(of approximately $8 \times 10^{22}$ aligned spins)
but a negligible magnetic moment.
The apparatus is shielded from external magnetic fields,
so any signal would represent a nonmagnetic interaction
coupling to the electron spins.
To search for a spin-monopole coupling,
the torsion pendulum is mounted on a turntable
that rotates about the suspension axis with angular frequency $\om$,
and a time-varying signal harmonically related to $\om$ is sought. 
Assuming an initial alignment of $\vec S$ along the $\hat x$ axis defined
in the laboratory frame,
the orientation of the spin vector $\vec S$ 
changes with the rotation as
\beq
\vec S = S(\cos \om t \, \hat x + \sin \om t \, \hat y)
\quad .
\label{S}
\eeq
This provides a second source of time dependence
for the potential $U$ in Eq.\ \rf{U}.

The potential $U$ induces 
a torque $\ta$ on the pendulum about the $\hat z$ axis.
The overall time dependence of the torque 
can be calculated from the potential $U$ 
as the cross product of 
the projection of the vector $\tilde b^{ej}$ onto the $x$-$y$ plane
with the spin vector $\vec S$.
The resulting expression in terms
of parameters in the nonrotating frame 
involves a sum of three harmonic terms
with angular frequencies $\om$ and $\om \pm \Om$: 
\bea
\ta
&=&  
2 \tilde b^e_Z S\sin \ch \sin \om t
+2\tilde b^e_\perp S\bigl\{
\sin^2\half\ch \sin [(\om-\Om)t+\be] 
\nonumber \\
&& 
\qquad\qquad
\qquad\qquad
-\cos^2\half\ch \sin [(\om+\Om)t-\be]
\bigr\} ,
\label{tau}
\eea
where 
$\tilde b^e_\perp
\equiv \sqrt{(\tilde b^e_X)^2 + (\tilde b^e_Y)^2}$
and $\be \equiv \tan^{-1} (\tilde b^e_Y/\tilde b^e_X)$.

The torque generates a pendulum twist angle $\th$ 
given by $\th = \ta/\ka$,
where $\ka$ is the pendulum spring constant.
As a function of time,
$\th$ can be obtained
as the solution of the differential equation
\beq
I \ddot \th + 2I\ga \dot \th + \ka \th = \ta
\quad ,
\label{diffeq}
\eeq
where $I$ is the moment of inertia
and $\ga$ is the damping constant
of the torsion pendulum.
Provided the rotational frequency $\om$ is much smaller than the
natural frequency $\om_0 = \sqrt{\ka/I}$,
oscillations with angular frequency $\om_0$ can be treated as irrelevant 
and the signal becomes the steady-state solution for $\th(t)$.
For the applied torque \rf{tau},
the steady-state solution is
\bea
\th(t)
&=& \fr{2S} \ka \biggl\{ 
\tilde b^e_Z {\cal A}_\om 
\sin \ch \sin \left(\om t - \de_{\om} \right)
\nonumber \\
&&  
+ \tilde b^e_\perp
\biggl[
{\cal A}_{\om - \Om} \sin^2\half\ch 
\sin [(\om-\Om)t-\de_{\om - \Om} + \be] 
\nonumber \\
&& 
-{\cal A}_{\om + \Om} \cos^2\half \ch  
\sin [(\om+\Om)t-\de_{\om + \Om}-\be]
\biggr]
\biggr\}
   \, ,
\label{theta}
\eea 
where 
${\cal A}_z \equiv \om_0^2 
\left[ (\om_0^2 - z^2)^2 + 4 \ga^2 z^2 \right]^{-1/2}$
is the attenuation factor and
$\de_z \equiv \tan^{-1} \left(2\ga z / (\om_0^2 - z^2) \right)$
is the phase shift due to the harmonic response of the pendulum
at frequency $z$.

The exact shape of $\th(t)$ is uncertain
because the relative sizes of the components 
$\tilde b^e_X$, $\tilde b^e_Y$, $\tilde b^e_Z$ are unknown.
However,
possible limiting cases can provide some insight.
Consider the E\"ot-Wash experiment, 
for which $\ch \simeq 42.3^o$.
If $\tilde b^e_Z \approx \tilde b^e_\perp$
then $\th(t)$ approximately vanishes
every sidereal period $T=2\pi / \Om$,
and $\th(t)$ oscillates at frequency $\om$ 
under an envelope with sidereal periodicity.
If instead $\tilde b^e_Z \ll \tilde b^e_\perp$
then the first term in $\th(t)$ is largely negligible,
and $\th(t)$ exhibits beats with approximate period $\half T$.
Finally,
if large $\tilde b^e_Z \gg \tilde b^e_\perp $
then the first term in $\th(t)$ dominates
and the sidereal variations disappear,
so $\th(t)$ merely oscillates with approximate frequency $\om$.

Given data taken with a rotating spin-polarized torsion pendulum of 
the E\"ot-Wash type, 
a test of Lorentz and CPT violation 
could proceed by extraction of the harmonic components 
with frequencies $\om$ and $\om \pm \Om$.
The amplitudes of these Fourier components would determine values
of all three parameters 
$\tilde b^e_X$, $\tilde b^e_Y$, and $\tilde b^e_Z$.
A compelling nonzero signal would provide 
evidence of Lorentz violation.
In the data analysis,
any summation or averaging process used to increase the statistics
would need to allow for the sidereal variation
to maintain the phases in the different terms in Eq.\ \rf{theta}.
The data already taken with the E\"ot-Wash II instrument 
are sensitive to the amplitude of twist-angle variations 
with frequency $\om$ at a level better than 10 nrads
\cite{harris98}.
If this accuracy can be achieved for all three Fourier components,
then impressive bounds of about $10^{-28}$ GeV 
could be attained on the components
$\tilde b^e_X$, $\tilde b^e_Y$, and $\tilde b^e_Z$.

In a search for spin-monopole couplings,
a preliminary analysis of data 
taken with the E\"ot-Wash II apparatus has been performed 
\cite{harris98}.
This analysis involves averaging results obtained 
at different sidereal times 
and extracting the amplitude of the harmonic components 
with frequencies equal to multiples of $\om$
(but not $\om \pm \Om$).
The averaging process maintains
the phase associated with the frequency $\om$ 
but not those associated with $\om\pm \Om$.
In the context of Eq.\ \rf{theta},
terms other than the first would therefore tend to average to zero
in the large-statistics limit
and so only the sensitivity to $\tilde b^e_Z$ remains.

The analysis yields the preliminary measurement
of a time-varying signal for $\th(t)$ with angular frequency $\om$ 
and amplitude $8.9 \pm 2.1 \pm 4.6$ nrads.
This time-varying signal provides a measurement of
$|\tilde b^e_Z| \simeq (1.4 \pm 0.8) \times 10^{-28}$ GeV,
where the two errors have been combined in quadrature.
Note that this value 
is almost an order of magnitude below 
the best bound on Lorentz violation in the electron sector
obtained to date in clock-comparison experiments
\cite{cctests,kl99a}.
Also,
the ratio
$r^e_{\rm spin} \equiv |\tilde b^e_Z|/m \simeq 3 \times 10^{-25}$
of this value to the electron mass compares favorably 
to the dimensionless suppression factor 
$m/M_{\rm Planck} \simeq 5 \times 10^{-20}$
that might be expected to govern 
spontaneous Lorentz and CPT breaking arising from the Planck scale
\cite{fn1}.
Confirmation that this preliminary result is a
signal for Lorentz and CPT violation
could emerge from a data reanalysis
extracting the amplitude of the harmonic components 
with frequencies $\om \pm \Om$ 
if nonzero amplitudes 
are detected in the ratio predicted by Eq.\ \rf{theta}.
This would also yield a measurement of $\tilde b^e_\perp$.

We conclude with some remarks about a different type of experiment 
using macroscopic spin-polarized matter,
in which the induced magnetization in a magnetic substance is studied. 
An experiment of this type has recently been performed at the
National Tsing Hua University in Taiwan
\cite{ni99}.
Small changes in the induced magnetization in 
a sample of paramagnetic TbF$_3$ salt are measured using a dc SQUID.
The salt is shielded in a field-free environment,
and a copper mass is rotated about it with frequency $f$.
The experiment searches for a time variation in the induced
magnetization with frequency $f$.

The apparatus functions as a magnetometer 
with exceptional sensitivity to an effective potential per volume 
\beq
u_{\rm eff} = \vec M \cdot \vec B_{\rm eff}
\quad 
\label{Ueff}
\eeq
for anomalous couplings between the magnetization $\vec M$ 
and an effective field $\vec B_{\rm eff}$.
With the correspondence
$(B_{\rm eff})_j = \tilde b^e_j/\mu_B$ in the laboratory frame,
where $\mu_B$ is the Bohr magneton,
the form of this potential matches that 
obtained for Lorentz and CPT violation
via Eq.\ \rf{h}.

Analysis of data taken with this apparatus provides
\cite{ni99}
an upper bound on $B_{\rm eff}$ of approximately $10^{-12}$ G.
This precision is achieved by accumulating large statistics,
which is made possible by 
rotating the copper mass around the salt crystal 
at the relatively high frequency of $f \simeq 0.96$ Hz.
However,
in the context of the standard-model extension,
no variation in the magnetization is caused by rotating a

A test of Lorentz and CPT symmetry with this apparatus
could nonetheless be performed
by searching for sidereal time variations
in the magnetization of the salt crystal.
An alternative possibility might be to rotate the entire apparatus 
on a turntable as in the E\"ot-Wash II instrument.
If the above sensitivity of $10^{-12}$ G
could be achieved for an effective field $(B_{\rm eff})_j$
due to Lorentz and CPT violation,
it would make attainable bounds on a combination of
$\tilde b^e_X$, $\tilde b^e_Y$, and $\tilde b^e_Z$ 
at the level of $10^{-29}$ GeV.
Another option for improving the sensitivity of
experiments of this type 
might be their inclusion in satellite-based tests of Lorentz symmetry,
perhaps in conjunction with a program for testing the equivalence principle
\cite{sat}.

\smallskip

We thank E.G.\ Adelberger and B.R.\ Heckel for discussions.
This work is supported in part by the U.S.\ D.O.E.\
under grant number DE-FG02-91ER40661 and by the N.S.F.\ 
under grant number PHY-9801869.

\end{multicols}
\end{document}